\documentclass[amsmath,amssymb,prb,
  twocolumn, floatfix, superscriptaddress,longbibliography]{revtex4-2}
  
\usepackage{graphicx}
\usepackage{amsmath}
\usepackage{amssymb}
\usepackage{mathrsfs}
\usepackage[T1]{fontenc}
\usepackage{calligra}
\usepackage{array}
\usepackage{float}
\usepackage{bm}
\usepackage[toc,page]{appendix}
\usepackage{braket}
\usepackage{ulem}
\usepackage[dvipsnames]{xcolor}
\usepackage{wrapfig}
\usepackage[colorlinks=true, linkcolor=blue, urlcolor=blue,
            citecolor=blue]{hyperref}

\begin{document}
	
	\title{Nonperturbative approach to interfacial spin-orbit torques induced by Rashba effect}
	
	\author{Alessandro Veneri}
	\affiliation{Department of Physics and York Centre for Quantum Technologies, University of York, YO10 5DD, York, United Kingdom}
	
	\author{David T. S. Perkins}
	\affiliation{Department of Physics and York Centre for Quantum Technologies, University of York, YO10 5DD, York, United Kingdom}
	
	\author{Aires Ferreira}
	\affiliation{Department of Physics and York Centre for Quantum Technologies, University of York, YO10 5DD, York, United Kingdom}
	
	\begin{abstract}
	    Current-induced spin-orbit torque (SOT) in normal metal/ferromagnet (NM/FM) bilayers bears great promise for technological applications, but the microscopic origin of purely interfacial SOTs  in ultra-thin systems is not yet fully understood. Here, we show that a linear response theory with a nonperturbative treatment of spin-dependent interactions and impurity scattering potential predicts damping-like SOTs that are strictly absent in perturbative approaches. The technique is applied to a two-dimensional Rashba-coupled ferromagnet (the paradigmatic model of a NM/FM interface), where higher-order scattering processes encoding skew scattering from nonmagnetic impurities allow for current-induced spin polarization with nonzero components along all spatial directions. This is in stark contrast to previous results of perturbative methods (neglecting skew scattering), which predict a coplanar spin-polarization locked perpendicular to the charge current as a result of conventional Rashba-Edelstein effect. Furthermore, the angular dependence of ensuing SOTs and their dependence upon the scattering potential strength is analysed numerically. Simple analytic expressions for the spin-density--charge-current response function, and related SOT efficiencies, are obtained in the weak scattering limit. We find that the extrinsic damping-like torques driven by impurity scattering reaches efficiencies of up to 7\% of the field-like (Rashba-Edelstein) torque. Our microscopic theory shows that bulk phenomena, such as the spin Hall effect, are not a necessity in the generation of the damping-like SOTs of the type observed in experiments on ultra-thin systems.
	\end{abstract}
	
	\maketitle
	
	\section{Introduction} \label{Sec_intro}
	
    The spin-orbit torque (SOT)~\cite{Manchon2019} is a phenomenon in which an unpolarized charge current injected into a normal metal/ferromagnetic metal (NM/FM) bilayer with inversion symmetry-breaking spin-orbit coupling (SOC) induces a nonequilibrium spin density, $\mathbf{S}$, in the NM and hence a torque, $\mathbf{T} \propto \mathbf{S} \times \mathbf{M}$. This torque then drives the dynamics of the FM layer's magnetization, $\mathbf{M}$, which can be switched by an electric current from one static configuration to another or enter into steady-state precession~\cite{ralph_spin_2008}. Such spin-torque-driven magnetization dynamics offers up a plethora of spintronic applications \cite{Locatelli2014,Ramaswamy2018,brataas_current-induced_2012,garello_ultrafast_2014}. In comparison to spin-transfer torque devices \cite{ralph_spin_2008}, SOT allows for faster and more energy-efficient devices \cite{cubukcu_spin-orbit_2014,ramaswamy_recent_2018,dolui_proximity_2020,liu_ultrafast_2010,ramaswamy_recent_2018}.
	
	Interfacial SOTs (i.e. those associated to purely interfacial effects) can arise from the lack of inversion symmetry in the stacking direction of NM/FM bilayers \cite{Manchon2019}, which yields charge-to-spin conversion processes resulting in the appearance of a spin accumulation in the NM at the interface --- a transport phenomenon commonly referred to as the Rashba-Edelstein effect (REE) \cite{edelstein_spin_1990,Silsbee_01,sanchez_spin--charge_2013}. In this scenario, a charge current passing through the NM layer generates a spin accumulation, $\mathbf{S}$, at the material's surface at bilayer's interface \cite{Inoue2003} which then exerts a torque on the magnetization of the FM partner due to the proximity coupling between itinerant electron spins and localized spins. Broadly speaking, two types of SOT can be generated: the first is called \textit{damping-like} (DL) SOT, which tends to align $\mathbf{M}$ with the effective magnetic field, $\mathbf{H}_{\text{eff}}$, acting upon the local local magnetic moments of the FM (this comprises the demagnetization field, anisotropy field and any applied external magnetic fields \cite{Manchon2019}). The second type of torque is called \textit{field-like} (FL) SOT, which causes $\mathbf{M}$ to precess about $\mathbf{H}_{\text{eff}}$. It is common practice in the literature to identify these torques by their odd/even nature in the magnetization, that is, $\textbf{T}_{\textrm{DL}}\equiv \textbf{T}_{\textrm{e}}$ and  $\textbf{T}_{\textrm{FL}}\equiv \textbf{T}_{\textrm{o}}$ with $\textbf{T}_{\textrm{e}}(\textbf{M})=\textbf{T}_{\textrm{e}}(-\textbf{M})$ and $\textbf{T}_{\textrm{o}}(\textbf{M})=-\textbf{T}_{\textrm{o}}(-\textbf{M})$. Whilst we have introduce this convention here, we would like to note that this is only strictly true for specific torque terms that appear to leading order in the magnetization, namely, the conventional FL SOT ($\textbf{T}_{\textrm{FL}}\propto  \textbf{M} \times \hat y$) and DL SOT ($\textbf{T}_{\textrm{DL}}\propto \textbf{M} \times \textbf{M} \times \hat y$, for a charge current applied along $\hat x$). However, a rigorous determination of the DL and FL torques can be achieved via a vector spherical harmonics expansion \cite{belashchenko_interfacial_2020}. For simplicity though, we shall use the common naming convention of the SOT types.
	
    In practice, there are two main mechanisms driving SOTs at NM/FM bilayers: the spin Hall effect (SHE) appearing in the bulk of the NM, and the REE appearing at the interface. A phenomenological study, along with a perturbative semiclassical Boltzmann analysis, of the SOTs generated by the SHE and REE was presented in Ref. \cite{amin_spin_2016}. In this work they observed that both the DL and FL torques stemming from the SHE became vanishingly small as the system thickness was decreased (the SHE is effectively suppressed when the spin diffusion length exceeds the NM thickness \cite{Zhang_00,Stamm_17}). In contrast, only the DL torque of the interfacial REE became negligible in the ultra-thin limit, whilst the REE's FL torque remained approximately constant.  However, several experiments on thin bilayers have observed torques that cannot be captured purely by the SHE, thus indicating the presence of significant interfacial DL torques \cite{garello_symmetry_2013,kim_layer_2013}. In fact, a study of an ultra-thin metallic bilayer (with a thickness below 1 nm), where SHE contributions are vanishingly small, still observed non-negligible DL torques responsible for the magnetic switching of the FM whose origin must be the interface \cite{kim_layer_2013}. Meanwhile, microscopic theories of interfacial SOTs have been put forward \cite{manchon_theory_2008,ado_microscopic_2017}, though have still failed to capture DL torques large enough to explain experimental observation in ultra-thin NM/FM bilayers \cite{kim_layer_2013}, as well as the  anisotropy of the DL torque \cite{garello_symmetry_2013}. Although these microscopic models handled the important role played by disorder in the NM, they did so within the Gaussian (white-noise) approximation where the scattering potential is treated perturbatively, whilst also handling the magnetic exchange interaction in a similar manner. The key finding of these early studies is the \textit{complete absence of DL torques} in the two-dimensional Rashba-coupled ferromagnet model, once vertex corrections due to impurity scattering are incorporated \cite{ado_microscopic_2017}.
	
	The use of perturbative methods has been questioned in a recent study \cite{sousa_skew-scattering-induced_2020}, where strong impurity scattering and the rich evolution of equilibrium spin textures with the Fermi level were seen to play a crucial role in the build up of nonequilibrium spin polarization and associated SOTs in  van der Waals heterostructures. Motivated by these developments, in this paper we shed light on the microscopic origin of interfacial SOTs in diffusive metallic bilayers and demonstrate that \textit{DL torques with nontrivial angular dependence} can be generated purely at the interface due to the interplay of Rashba SOC, magnetic proximity effect and impurity scattering. To this end, we formulate a  linear response theory that is nonperturbative in both the impurity scattering strength and spin interactions (magnetic exchange and Rashba SOC) to calculate the current-induced spin polarization in the NM, $\mathbf{S}$.  We achieve this by modelling the NM as a two-dimensional electron gas (2DEG) and employ a generalized self-consistent diagrammatic technique that handles disorder at the complete $T$-matrix level \cite{milletari_quantum_2016,milletari_covariant_2017,offidani_optimal_2017,offidani_microscopic_2018,offidani_anomalous_2018} to calculate the spin-density--charge-current response functions, whilst allowing for the FM's magnetization to lie at an arbitrary angle. The anisotropic spin texture of 2DEG's Fermi rings can be seen to enrich the possible current-induced spin polarizations (with hitherto unseen  non-zero components along all principal axes emerging as a result of skew scattering of conduction electrons) and hence we predict new types of interfacial SOT with extrinsic origin. We present analytical results in the  weak scattering limit (incorporating skew scattering), and use a numerical procedure to extract the full angular dependence of the SOTs.
	
	The remainder of this paper is structured as follows: In Sec. \ref{Sec_theory}, we formally introduce current-induced SOTs exerted by the NM and the Hamiltonian for disordered 2DEGs with symmetry breaking SOC. We then present a self-consistent diagrammatic theory to evaluate the disorder-averaged linear response, by making use of the $T$-matrix approach. Afterwards, in Sec. \ref{Sec_Results}, we provide an intuitive semiclassical picture for the SOT in NM/FM bilayers before then applying the diagrammatic method to 2DEGs in the weak  scattering scattering limit. Additionally, we provide a numerical study of the SOT in the strong scattering limit as well as the dependence of the SOT coefficients upon the magnetization's orientation. In Sec. \ref{Sec_conclusions} we present our conclusions.

	\section{Models and Methods} \label{Sec_theory}
	
	\subsection{SOT components and notation} \label{Subsec_SOT_dynamics}
	The dynamical effects of SOT can be modelled by including the additive term \cite{Manchon2019}
	\begin{equation}
		\mathbf{T} = \frac{\gamma}{t\,M_{s}} \mathbf{H}_{\text{SOT}} \times \mathbf{m},
		\label{SOT_term}
	\end{equation}
	into the Landau-Lifshitz-Gilbert equation \cite{brataas_current-induced_2012}, where $t$ is the FM's thickness, $\mathbf{m} = \frac{\mathbf{M}}{|\mathbf{M}|}$, $\gamma$ is the gyromagnetic ratio, $M_{s}$ the saturation magnetization of the magnetic layer, and $\mathbf{H}_{\text{SOT}}$ is the effective magnetic field [Eq. (\ref{SOT_field})] generated by the nonequilibrium spin polarization of conduction electrons. Within the theory of linear response, we may write the spin--orbit effective field as
	\begin{equation}
		\mathbf{H}_{\text{SOT}} = \Delta_{\text{xc}} \hat{K}_{J} \, \mathbf{J},
		\label{SOT_field}
	\end{equation}
	where $\Delta_{\text{xc}}$  is the interfacial exchange coupling, $\mathbf{J}=\hat \sigma \textbf{E}$ is the in-plane electric current density, $\textbf{E}$ is the external electric field and $\hat{K}_{J}$ is related to the spin susceptibility, $\hat{K}$, and electrical conductivity tensor, $\hat{\sigma}$, by $\hat K_{J} = \hat K \cdot \hat\sigma^{-1}$. 
  	
	To separate the torque term in Eq. (\ref{SOT_term}) into damping-like (even in $\mathbf{m}$) and field-like (odd in $\mathbf{m}$) parts, $\mathbf{T}^{\text{e}}$ and $\mathbf{T}^{\text{o}}$ respectively, we perform a symmetry analysis based on  the following decomposition  
	\begin{subequations}
	\begin{equation}
		\mathbf{T}^{\text{e}} = \frac{\Delta_{\text{xc}}\gamma}{t\,M_{s}} (\tau_{1}^{\text{e}} \mathbf{m} \times (\mathbf{m} \times (\mathbf{e}_{z} \times \mathbf{J})) + \tau_{2}^{\text{e}} \mathbf{m} \times \mathbf{e}_{z} (\mathbf{m} \cdot \mathbf{J})),
	\end{equation}
	\begin{equation}
		\mathbf{T}^{\text{o}} = \frac{\Delta_{\text{xc}}\gamma}{t\,M_{s}} (\tau_{1}^{\text{o}} \mathbf{m} \times (\mathbf{e}_{z} \times \mathbf{J}) + \tau_{2}^{\text{o}} \mathbf{m} \times (\mathbf{m} \times \mathbf{e}_{z}) (\mathbf{m} \cdot \mathbf{J})),
	\end{equation}
	\end{subequations}
    which assumes an interface with continuous rotational symmetry about the $z$-axis \cite{miron_perpendicular_2011,skinner_spin-orbit_2014,wang_diffusive_2012,haney_current_2013}. The torque efficiencies, $\tau_{i}^{j}$, are the controlling parameters of SOT and hence are the primary focus of our work. The torque efficiencies may be written in terms of the magnetization and spin susceptibility tensor components as
	\begin{subequations}
	\begin{equation}
		\tau_{1}^{\text{e}} = \frac{ K_{yy}}{m_{z}},
	\end{equation}
	\begin{equation}
		\tau_{2}^{\text{e}} = \ \frac{K_{xx} - K_{yy}}{m_{x}^{2}m_{z}} - \frac{K_{xx}}{m_{z}} - \frac{K_{zx}}{m_{x}},
	\end{equation}
	\begin{equation}
		\tau_{1}^{\text{o}} =  K_{xy} - \frac{m_{x}}{m_{z}} K_{zy},
	\end{equation}
	\begin{equation}
		\tau_{2}^{\text{o}} = \frac{1}{m_{x}} \left( \frac{K_{xy} +K_{yx}}{m_{x}} - \frac{K_{zy}}{m_{z}} \right).
	\end{equation}
	\label{Torque_efficiencies_general}%
	\end{subequations}
	From these expressions, we clearly see that $K_{xx}$, $K_{yy}$, and $K_{zx}$ contribute solely to the damping-like torque, whilst $K_{xy}$, $K_{yx}$, and $K_{zy}$ generate the field-like torque.
	
	\subsection{The Hamiltonian} \label{Subsec_Hamiltonian}
	The general Hamiltonian for the FM partner material may be written as
	\begin{equation}
		\mathcal{H} = \mathcal{H}_{0} + \mathcal{H}_{\text{PE}} + \mathcal{H}_{\text{dis}},
	\end{equation}
	where $\mathcal{H}_{0}$ is the Hamiltonian for the clean isolated NM, $\mathcal{H}_{\text{PE}}$ accounts for proximity effects, and $\mathcal{H}_{\text{dis}}$ describes the impurity landscape. In this paper we  model the NM as a 2DEG, whilst focusing on the effects of Rashba SOC and the exchange interaction between classical magnetic moments in the FM and the spin of conduction electrons in the NM. We may therefore write $\mathcal{H_{\text{PE}}} = \mathcal{H}_{\text{BR}}+\mathcal{H}_{\text{xc}}$, where the first and second terms correspond to Rashba SOC and the exchange interaction respectively. To describe the spin-dependent interactions in a compact form, we introduce the non-Abelian SU(2) gauge field \cite{Tokatly_Sherman_10}
	\begin{equation}
		\mathcal{A}^{\mu} = \mathcal{A}_{i}^{\mu} s_{i} \qquad (i=0,x,y,z),
	\end{equation}
	with $s_i$ ($i=x,y,z$) are Pauli matrices acting in spin space and $s_0$ is the identity matrix. This field is then inserted into $\mathcal{H}_{0}$ as a generalized vector potential in an analogy to minimal coupling.  For a standard Rashba-coupled ferromagnet, we have only the non-zero components being
	\begin{equation}
		\mathcal{A}_{x}^{y} = -\mathcal{A}_{y}^{x} = \alpha m^{*}, \quad \mathcal{A}^{0} = - \Delta_{\text{xc}} \mathbf{m} \cdot \mathbf{s} \equiv -\boldsymbol{\Delta} \cdot \mathbf{s},
	\end{equation}
	where $m^{*}$ is the effective electron mass, $\alpha$ is the SOC strength, and $\Delta_{\text{xc}}$ is the exchange coupling (here we assumed a conventional isotropic Zeeman interaction \cite{Anderson1970}).
	
	In the absence of disorder, the NM Hamiltonian takes the second-quantized form
	\begin{equation}
		 \mathcal{H}_{0} + \mathcal{H}_{\text{PE}}  = \int d\mathbf{x} \, \psi^{\dagger}(\mathbf{x}) \left[ \frac{(\mathbf{p}+\boldsymbol{\mathcal{A}})^{2}-\boldsymbol{\mathcal{A}}^{2}}{2m^{*}} - \varepsilon - \mathcal{A}^{0} \right] \psi(\mathbf{x}).
		\label{2DEG_Hamiltonian}
	\end{equation}
	Here $\varepsilon$ is the Fermi energy, $p^{\mu} = (-\varepsilon/v,\mathbf{p})$ is the 3-momentum operator, and $v$ is  the Fermi velocity of the electrons. The disorder term in the Hamiltonian, $\mathcal{H}_{\text{dis}}$, reads as
	\begin{subequations}
	\begin{equation}
		\mathcal{H}_{\text{dis}}=\int d\mathbf{x} \, \psi^{\dagger}(\mathbf{x}) \, V(\mathbf{x}) \, \psi(\mathbf{x}),
	\end{equation}
	\begin{equation}
		V(\mathbf{x}) = \sum_{i} W(\mathbf{x}-\mathbf{x}_{i}),
	\end{equation}
	\end{subequations}
	where $V(\mathbf{x})$ is the total impurity potential, and $W(\mathbf{x}-\mathbf{x}_{i})$ is the potential of a single impurity located at position $\mathbf{x}_{i}$ within the NM/FM interface (the areal density of impurities is denoted as $n$). We note that in general, $V(\mathbf{x})$ has a matrix structure and can include effects such as local SOC and magnetic impurities \cite{huang_spin-charge_2017}, but these are not a necessity in the generation of FL and DL SOTs as we shall show below. To recover macroscopic results, we average over all possible impurity configurations within the $T$-matrix formalism \cite{milletari_quantum_2016}, which accounts for all possible scattering scenarios involving a single impurity. In this paper we work in the dilute limit ($n \ll 1$) with a focus on short-range scalar impurities: $W(\mathbf{r}) = u R^{2} \delta(\mathbf{r})$ where $u$ is the scattering potential, $R$ is the characteristic length scale of the impurity potential's range and $\delta(\mathbf{r})$ is the delta (Dirac) function.
	
	Having set up the Hamiltonian describing the NM, our focus now turns to calculating the generalized spin-density--charge-current response tensor. The next section details our theory for obtaining response functions without the need for a perturbative treatment of the exchange coupling, Rashba SOC or impurity scattering potential. The only perturbative parameter governing the validity of our diagrammatic theory is $(\varepsilon \tau)^{-1} \ll 1$, where $\tau$ is the momentum relaxation time.
	
	\begin{figure}
	\centering
		\includegraphics[width=1\linewidth]{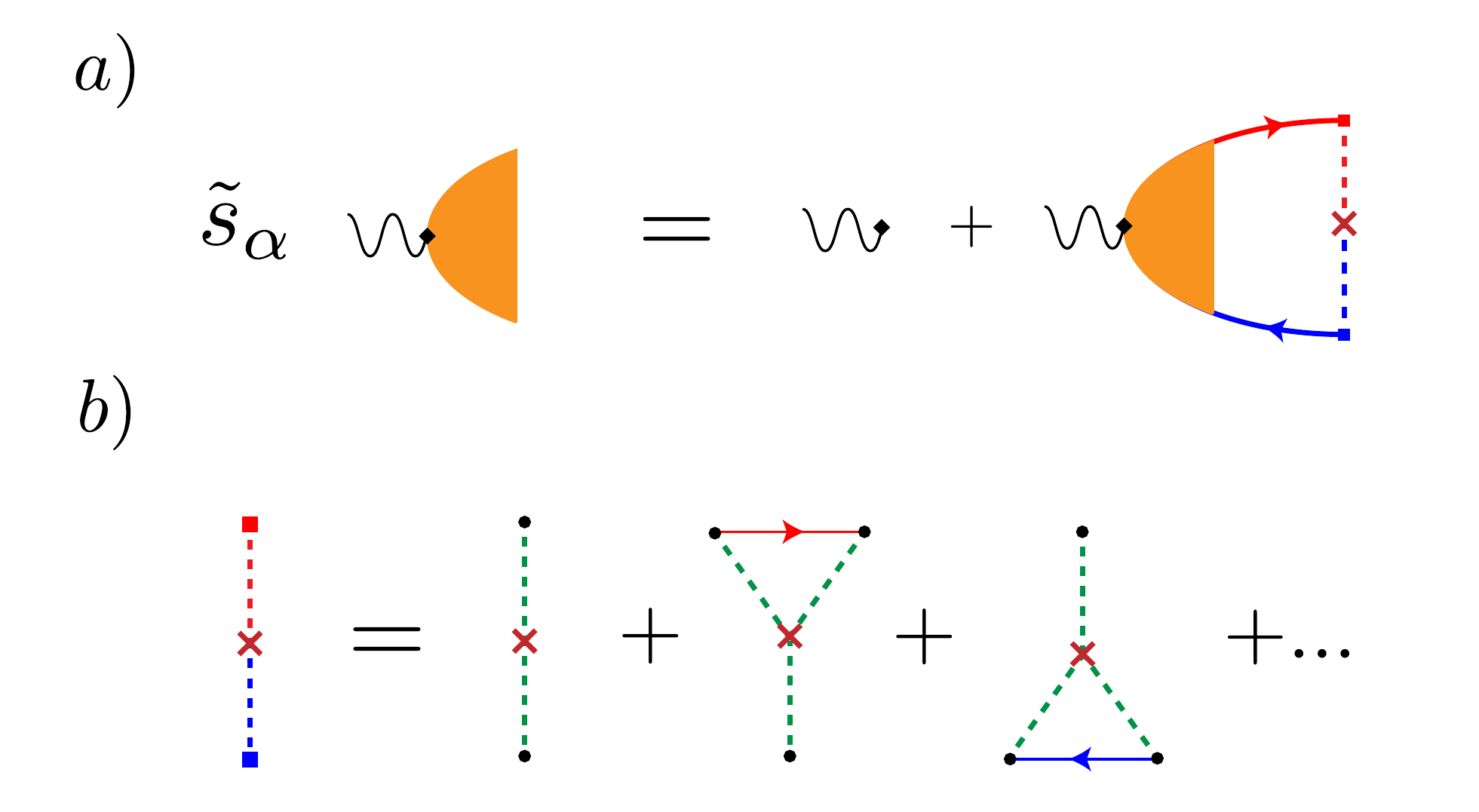}
		\caption{Diagrammatic expansion of the zero temperature spin-density--charge-current response function: (a) the disorder-renormalized spin density vertex function and (b) the $T$-matrix skeleton expansion. Solid lines with arrows denote disorder averaged Green's functions, while green dashed lines represent single impurity potential insertions.  Red/blue indicate advanced/retarded sectors.}
		\label{Diagrammatics}
	\end{figure}
	
	\subsection{Diagrammatic theory and the T-matrix} \label{Subsec_diagrammatics}
	To understand a system's response to spin-charge conversion away from equilibrium, we employ the theory of linear response. In particular, we assume that $\mathbf{M}$ and $\mathbf{E}$ vary slowly in both position and time  (i.e. on scales larger than the mean free path and $\tau$), and hence neglect their spatial and temporal dependence. The response of the spin density to the electric field is then simply (assuming Einstein summation)
	\begin{equation}
		S_{\alpha} = K_{\alpha\beta}E_{\beta},
		\label{LRT}
	\end{equation}
	where $K_{\alpha\beta}$ ($\alpha=x,y,z$ and $\beta=x,y$) is the spin susceptibility response tensor  with a $3\times2$ matrix structure in our case (c.f. Eq. (\ref{SOT_field})). Therefore, the effect of SOT upon the FM is contained entirely within the object $K_{\alpha\beta}$, which we shall treat using the Kubo-Streda formula \cite{crepieux_theory_2001}. This spin-current response function can be separated into two contributions,
	\begin{equation}
		K_{\alpha\beta} = R_{\alpha\beta}^{0} + R_{\alpha\beta}^{\text{\ensuremath{\varepsilon}}},
	\end{equation}
	where $R_{\alpha\beta}^{0}$ is the system's Fermi sea (type II) response, and $R_{\alpha\beta}^{\varepsilon}$ is the Fermi surface (type I) contribution to the total response. Written explicitly, the Fermi surface response takes the form
	\begin{equation}
		R_{\alpha\beta}^{\epsilon} = -\frac{1}{4\pi} \langle \text{Tr}[(s_{\alpha} G^{+} j_{\beta} - j_{\beta} G^{-} s_{\alpha}) (G^{+}-G^{-})]\rangle_{\text{dis}},
		\label{Generic_Kubo}
	\end{equation}
	where $j_{\beta} = e \, \partial\mathcal{H}/\partial p_{\beta}$ is the electric current operator \cite{rashba_spin_2003} ($e<0$), $\langle...\rangle_{\text{dis}}$ denotes disorder averaging, $G^{\pm}=(\epsilon-\mathcal{H} \pm i\delta)^{-1}$ is the clean retarded$(+)$/advanced$(-)$ Green's function at the Fermi surface, and $\delta$ is a positive infinitesimal. By working in the dilute limit (i.e. low impurity concentration) we may neglect the Fermi sea contribution, and ignore terms containing products of the same Green's function in Eq. (\ref{Generic_Kubo}) \cite{milletari_quantum_2016},
	\begin{equation}
		K_{\alpha\beta} \simeq \frac{1}{2\pi} \text{Tr}[\langle s_{\alpha} G^{+} j_{\beta} G^{-} \rangle_{\text{dis}}].
		\label{Disorder_averaged_response_function_explicit}
	\end{equation}
	
	Applying the disorder average yields
	\begin{equation}
		K_{\alpha\beta} = \frac{1}{2\pi}\sum_{\mathbf{p}}\text{tr}[\tilde{s}_{\alpha} \mathcal{G}^{+}_{\mathbf{p}} j_{\beta} \mathcal{G}^{-}_{\mathbf{p}}],
		\label{Disorder_averaged_response_function_simplified}
	\end{equation}
	where we have written the response function in momentum space explicitly and the trace is now over the internal matrix indices. We perform our calculations using the standard rules of diagrammatics and assume the non-crossing approximation \cite{Rammer_book}. The retarded and advanced disorder-averaged Green's functions appearing in Eq. (\ref{Disorder_averaged_response_function_simplified}) are given by
	\begin{equation}
		\mathcal{G}^{\pm}_{\mathbf{p}} = \frac{1}{(G^{\pm}_{0,\mathbf{p}})^{-1}-\Sigma^{\pm}},
		\label{Disorder_averaged_GF_definition}
	\end{equation}
	where $G^{\pm}_{0,\mathbf{p}}$ is the clean Green's function, and $\Sigma^{\pm}$ is the retarded/advanced disorder self-energy. To leading order in the impurity density, $n$, we may relate $\Sigma^{\pm}$ to the $T$-matrix via $\Sigma^{\pm} = nT^{\pm}$. The advanced/retarded $T$-matrix is represented diagrammatically in Fig. \ref{Diagrammatics}, which can be shown to yield
	\begin{subequations}
	\begin{equation}
		T^{\pm} = \widetilde{W} \frac{1}{1 - \widetilde{W} g_{0}^{\pm}},
		\label{T_matrix_compact}
	\end{equation}
	\begin{equation}
		g_{0}^{\pm} = \int \frac{d^{2}p}{(2\pi)^{2}} G^{\pm}_{0,\mathbf{p}},
	\end{equation}
	\end{subequations}
	where $g_{0}^{\pm}$ is the momentum integrated clean Green's function, and $\widetilde{W}$ is the Fourier transfoorm of $W(\mathbf{r})$. Finally, the renormalized vertex (Fig. \ref{Diagrammatics}), $\tilde{s}_{\alpha}$, is given by the Bethe-Salpeter equation \cite{offidani_microscopic_2018-1},
	\begin{equation}
		\tilde{s}_{\alpha} = s_{\alpha} + n \sum_{\mathbf{p}} T^{-} \mathcal{G}^{-}_{\mathbf{p}} \tilde{s}_{\alpha} \mathcal{G}^{+}_{\mathbf{p}} T^{+}.
		\label{Renormalised_vertex_BS}
	\end{equation}
	
	Within the first Born approximation (FBA), Eq. (\ref{T_matrix_compact}) is expanded to second order in $W$. This leads to the renormalized vertex being given by a simple ladder series of impurity scatterings, and hence fails to capture the physics of skew scattering. This is represented by the first term in Fig. \ref{Diagrammatics}b. We must therefore perform a $T$-matrix expansion that is non-perturbative in the scattering potential \cite{hirschfeld_consequences_1988,ferreira_extrinsic_2014}.
	
	A common approximation accompanying the FBA is the Gaussian approximation, in which off-diagonal elements in the FBA (recall we are working with $2\times2$ spin matrices) are also neglected. However, we note that, even with scalar impurities, Eq. (\ref{T_matrix_compact}) allows for the $T$-matrix to possess non-zero off-diagonal elements. Therefore, in order to create a fully self-consistent theory, we must include these elements and so the Gaussian approximation is not appropriate for understanding the non-equilibrium spin density induced in the NM.
	
	To account for all terms and the matrix structure of $T^{\pm}$, we simplify the renormalized vertex by projecting Eq. (\ref{Renormalised_vertex_BS}) onto the Pauli algebra (here $\alpha,\beta,\nu \in \{0,x,y,z\}$), 
	\begin{equation}
		\tilde{s}_{\alpha} = \mathcal{D}_{\alpha\beta} s_{\beta}, \qquad \mathcal{D}_{\alpha\beta} = \frac{1}{2} \text{tr}[\tilde{s}_{\alpha}s_{\beta}].
	\end{equation}
    The coefficients are then given by
	\begin{equation}
	\begin{split}
		\mathcal{D}_{\alpha\beta} &= \delta_{\alpha\beta} + \mathcal{D}_{\alpha\nu} \mathcal{M}_{\nu\beta}, \\
		\mathcal{M}_{\nu\beta} &= \frac{n}{2} \sum_{\mathbf{p}} \text{tr}[T^{-} \mathcal{G}^{-}_{\mathbf{p}} s_{\nu} \mathcal{G}^{+}_{\mathbf{p}} T^{+} s_{\beta}].
	\end{split}
	\end{equation}
	To evaluate the $\mathcal{M}$ matrix, we decompose it into two separate matrices,
	\begin{equation}
	\begin{split}
		\Upsilon_{\alpha\beta} &= \frac{1}{2} \text{tr}[s_{\alpha}T^{+}s_{\beta}T^{-}], \\
		\mathcal{N}_{\alpha\beta} &= \frac{n}{2} \sum_{\mathbf{p}} \text{tr}[s_{\alpha} \mathcal{G}^{+}_{\mathbf{p}} s_{\beta} \mathcal{G}^{-}_{\mathbf{p}}],
	\end{split}
	\end{equation}
	such that $\mathcal{M} = \mathcal{N} \Upsilon$. The $\Upsilon$ matrix describes the insertion of impurities connecting the two sides of the response bubble, whilst the $\mathcal{N}$ matrix encodes information about the disorder-averaged Green's functions forming a response bubble in the absence of interference.
	
	The projection coefficients are in fact the elements of the \textit{generalized Diffuson} operator
	\begin{equation}
		\mathcal{D} = (1 - \mathcal{M})^{-1}.
	\end{equation}
	Consequently Eq. (\ref{Disorder_averaged_response_function_simplified}) becomes
	\begin{equation}
		K_{\alpha\beta} = \frac{1}{2\pi} \sum_{\mathbf{p}} \text{tr}[\mathcal{D}_{\alpha\nu} s_{\nu} \mathcal{G}^{+}_{\mathbf{p}} j_{\beta} \mathcal{G}^{-}_{\mathbf{p}}].
		\label{Disorder_averaged_response_function_Diffuson_general}
	\end{equation}
	Under the Gaussian approximation, $\Upsilon$ becomes the sum over all forms of scalar disorder, reducing the \textit{generalized Diffuson} to the standard form in the literature \cite{burkov_theory_2004}.
	
	Finally, in order to find $\hat{K}_{J}$ we need to perform an analogous treatment of the current-current response function to obtain the charge conductivity tensor. In this case, the response function is given by Eq.(\ref{Disorder_averaged_response_function_explicit}) with $s_{\alpha} \rightarrow j_{\alpha}$. Upon disorder averaging, we now choose to renormalize the $j_{\beta}$ vertex to yield
	\begin{equation}
		\sigma_{\alpha\beta} = \frac{1}{2\pi}\sum_{\mathbf{p}}\text{tr}[j_{\alpha} \mathcal{G}^{+}_{\mathbf{p}} \tilde{j}_{\beta} \mathcal{G}^{-}_{\mathbf{p}}].
		\label{Disorder_averaged_conductivity_simplified}
	\end{equation}
	The renormalized current vertex is found by letting $\tilde{j}_{\beta} = j_{\beta} + \delta j_{\beta}$, and then solving a Bethe-Salpeter equation for the corrections to the bare current vertex,
	\begin{subequations}
	\begin{equation}
		\delta j_{\beta} = \delta\bar{j}_{\beta} + n\sum_{\mathbf{p}} T^{-}\mathcal{G}^{-}_{\mathbf{p}} \delta j_{\beta} \mathcal{G}^{+}_{\mathbf{p}} T^{+},
	\end{equation}
	\begin{equation}
		\delta\bar{j}_{\beta} = n\sum_{\mathbf{p}} T^{-}\mathcal{G}^{-}_{\mathbf{p}} j_{\beta} \mathcal{G}^{+}_{\mathbf{p}} T^{+}.
	\end{equation}
	\end{subequations}
	Since we are working with the current vertex explicitly, we use the projection
	\begin{equation}
		\delta j_{\beta} = \widetilde{\mathcal{D}}_{\beta\nu} \delta \bar{j}_{\nu}, \qquad \widetilde{\mathcal{D}}_{\beta\nu} = \frac{1}{2}\text{tr}[\delta j_{\beta} \delta \bar{j}_{\nu}],
	\end{equation}
	and find an operator analogous to the Diffuson,
	\begin{equation}
		\widetilde{D} = (1 - \mathcal{N}^{\text{T}} \Upsilon^{\text{T}})^{-1}.
	\end{equation}
	The electrical conductivity tensor is thus given by
	\begin{equation}
		\sigma_{\alpha\beta} = \frac{1}{2\pi}\sum_{\mathbf{p}}\text{tr}[j_{\alpha} \mathcal{G}^{+}_{\mathbf{p}} \widetilde{\mathcal{D}}_{\beta\nu} j_{\nu} \mathcal{G}^{-}_{\mathbf{p}}].
		\label{Disorder_averaged_conductivity_Diffuson_general}
	\end{equation}
	
	Given the expressions for the spin susceptibility and electrical conductivity in Eq. (\ref{Disorder_averaged_response_function_Diffuson_general}) and Eq. (\ref{Disorder_averaged_conductivity_Diffuson_general}) respectively, we may compute the spin density response to the application of a charge current and hence calculate the effective magnetic field induced by the nonequilibrium spin density. This therefore allows us to determine the explicit form of the SOT [Eq. (\ref{SOT_field})] driving the magnetization dynamics of the FM. The efficiency of this charge-to-spin conversion (CSC) process can be defined as $\theta_{\alpha\beta} = -2ev K_{J,\alpha\beta}$ \cite{mellnik_spin-transfer_2014}.
	
	\begin{figure}[t]
		\centering
		\includegraphics[width=0.9\linewidth]{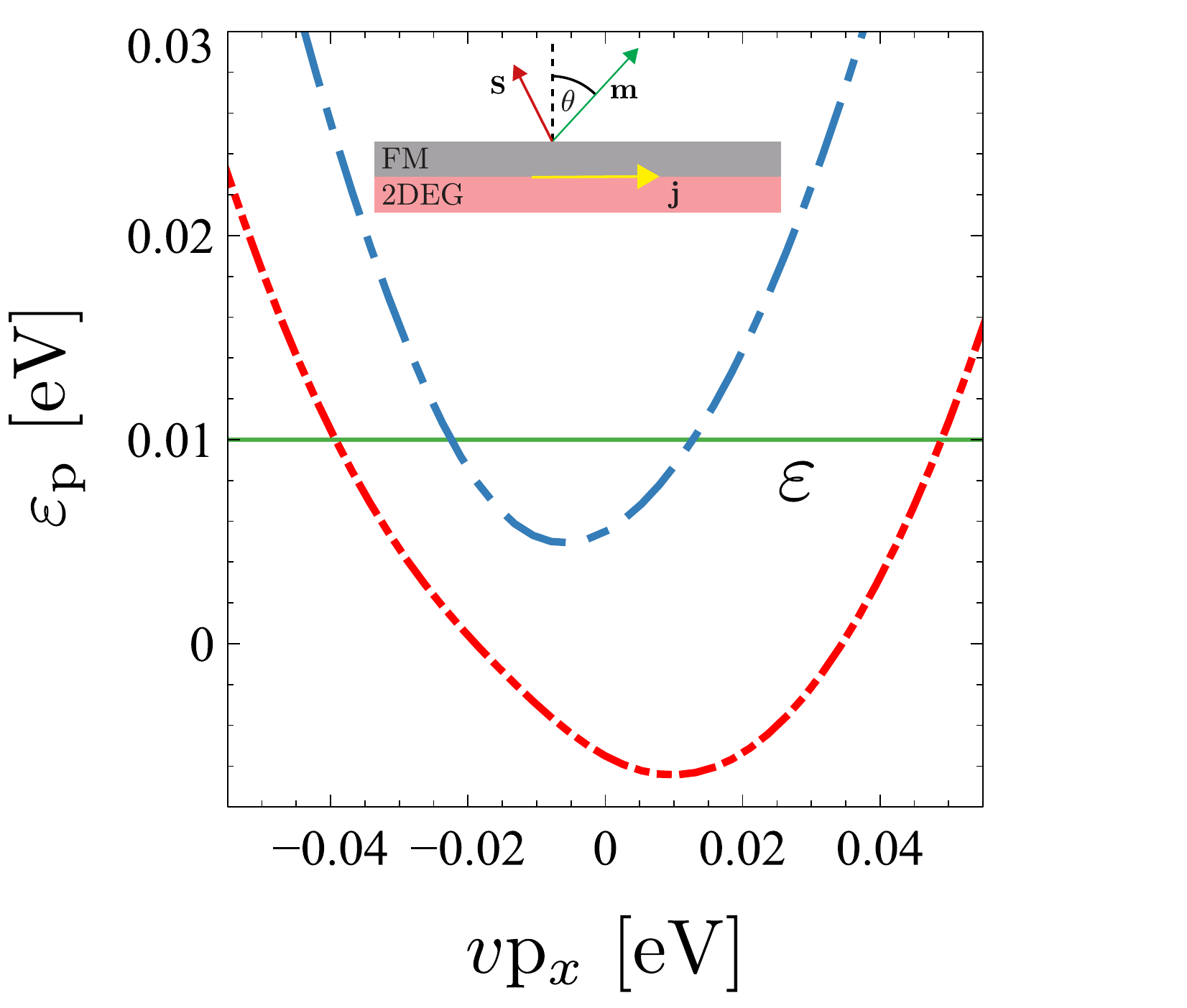}
		\caption{Band structure of a 2DEG with Rashba SOC and an exchange interaction, where we have assumed $m_{y} = 0$ without loss of generality, which shifts the Fermi rings along the $y$-axis by the $m_{x}$ component. Parameters: $\alpha = 1.7 \times 10^{-11}$ eVm, $\Delta_{\text{xc}} = 5.5$ meV, $v = 10^{5}\text{ ms}^{-1}$, and $\theta = \pi/4$. The green line represents the Fermi energy $\varepsilon = 0.01$ eV. Inset: NM/FM bilayer schematic, with electric current, $\mathbf{j}$, aligned with the $x$-axis, and FM magnetization at angle $\theta$ to the $z$-axis.}
		\label{2DEG_dispersion}
	\end{figure}
	
	\section{Results and Discussion} \label{Sec_Results}
	
	Starting from the disorder-free electron picture of a 2DEG, Rashba SOC causes spin-splitting of the parabolic dispersion into two bands with the electrons' spin being locked in-plane and perpendicular to their momentum. The spins of the upper and lower bands ($\nu$ is the band index) wind in a clockwise and anti-clockwise manner, respectively, around their corresponding Fermi rings. Next, the out-of-plane exchange interaction due to $m_{z}$ opens up a gap between the two bands, and leads to an out-of-plane tilting in each band's spin texture: the upper band spins rotate towards $m_{z}\mathbf{e}_{z}$, and the lower band spins rotate towards $-m_{z}\mathbf{e}_{z}$. Finally, the in-plane magnetization deforms the shape of the bands, whilst also shifting them in opposite directions along the axis perpendicular to the in-plane component. This generates a highly anisotropic dispersion relation, see Fig. \ref{2DEG_dispersion}, and so requires expansion in $m_{x}$ and $m_{y}$ to allow for analytic evaluation. For ease of reference going forward, we write the magnetization as
	\begin{equation}
		\mathbf{m} = \sin\theta \, \mathbf{e}_{x} + \cos\theta \, \mathbf{e}_{z},
	\end{equation}
	where we have assumed $m_{y} = 0$ without loss of generality.
	
	Applying an electric field, $\mathbf{E}$, to this system shifts the Fermi rings in the direction of $-\mathbf{E}$, leading to a non-zero centre-of-mass momentum and hence an electrical current. This shifting of the Fermi rings leads to an out-of-equilibrium spin accumulation due to the momentum dependence of the Fermi ring spin texture, $\mathbf{s}_{\mathbf{k}\nu}$. This result emerges naturally from the average of $\mathbf{s}_{\mathbf{k}\nu}$ away from equilibrium when considered in a semiclassical manner. To demonstrate this, let us consider the effect of small external perturbations upon the electron distribution function. The distribution function may then be written as $f_{\mathbf{k}\nu} + \delta f_{\mathbf{k}\nu}$, where $f_{\mathbf{k}\nu}$ is the Fermi function for the band $\nu$,
	\begin{equation}
		\delta f_{\mathbf{k}\nu} \propto |\mathbf{E}| \sum_{l} \left[ \tau_{\parallel}^{l,\nu} \cos(l\phi) + \tau_{\perp}^{l,\nu} \sin(l\phi) \right],
	\end{equation}
	is the linear correction to $f_{\mathbf{p}\nu}$ due to the external electric field, and $\phi$ is the azimuthal angle of the momentum. Note that the coefficients, $\tau_{\parallel}^{l,\nu}$ and $\tau_{\perp}^{l,\nu}$ are functions of $\alpha$, $m_{z}^{2}$, and $|\mathbf{k}|$. The spin polarization is then given by
	\begin{equation}
		\mathbf{S} = \sum_{\mathbf{k},\nu} \mathbf{s}_{\mathbf{k}\nu} \, \delta f_{\mathbf{k}\nu}.
	\end{equation}

	For a small in-plane magnetization the 2DEG's spin texture may be written as
	\begin{subequations}
	\begin{equation}
		\mathbf{s}_{\mathbf{k}\nu} = \nu(\mathbf{s}_{\mathbf{k}\nu}^{0} + m_{x} \, \delta\mathbf{s}_{\mathbf{k}\nu}),
		\label{Spin_texture_full_semiclassical_2DEG}
	\end{equation}
	\begin{equation}
		\mathbf{s}_{\mathbf{k}\nu}^{0} = \rho_{\parallel} \hat{\mathbf{k}} \times \mathbf{e}_{z} + \rho_{\perp} \mathbf{e}_{z},
		\label{Spin_texture_ESA_semiclassical_2DEG}
	\end{equation}
	\begin{equation}
	\begin{split}
		\delta\mathbf{s}_{\mathbf{k}\nu} = [\omega_{\parallel} &+ \xi_{\parallel} \cos(2\phi)] \mathbf{e}_{x} \\
		&+ \xi_{\parallel} \, \sin(2\phi) \, \mathbf{e}_{y} - \omega_{\perp} m_{z} \sin \phi \mathbf{e}_{z},
		\label{Spin_texture_correction_semiclassical_2DEG}
	\end{split}
	\end{equation}
	\label{Spin_texture_semiclassical}
	\end{subequations}where $\omega_{\parallel}$, $\omega_{\perp}$, and $\xi_{\parallel}$ are also functions of $\alpha$, $m_{z}^{2}$, and $|\mathbf{k}|$. The $\mathbf{s}_{\mathbf{k}\nu}^{0}$ term is responsible for the spin-helical part of $\mathbf{s}_{\mathbf{k}\nu}$, and therefore produces an imbalance in oppositely aligned spins that is transverse to the applied electric field. Clearly, this term is the origin of the familiar REE, depending entirely on Rashba SOC, and generates non-zero contributions to $K_{xy}$ and $K_{yx}$. These components survive the restrictions enforced by the FBA, where they appear independent of the magnetization.
	
	We can easily see from Eq. (\ref{Spin_texture_semiclassical}) that the presence of an in-plane magnetization allows for an angle-dependent out-of-plane spin accumulation. Hence, this correction contributes to the $K_{zx}$ and $K_{zy}$ elements. However, under the FBA we find that no such response is seen in the out-of-plane polarization, $S_{z}$, when $\varepsilon > \Delta_{\text{xc}}$, suggesting that the physics governing out-of-plane polarization is more sensitive to the scattering strength than REE. It turns out that the non-zero spin polarizations of the individual bands cancel out perfectly within the FBA, which explains the vanishing of $K_{zy}$ reported in Ref. \cite{ado_microscopic_2017}. Overall, there are 4 vanishing responses within the FBA, namely, $K_{zx}=K_{zy}=K_{xx}=K_{yy}=0$.
	
	To overcome the limitations of the FBA, we work with the scattering strength non-perturbatively by using the $T$-matrix approach detailed in section \ref{Subsec_diagrammatics}; this allows for skew scattering when the partner FM has a finite out-of-plane magnetization component, i.e. $m_{z} \neq 0$. As a result, we find that the $(k_x,k_y)$ and $(k_x,-k_y)$ points lying on the Fermi rings become inequivalent with different occupation numbers, and hence generate a non-zero value for $S_{z}$. The same mechanism generates the diagonal contributions, $K_{xx(yy)}$, which depends on $m_{z}$ but to leading order is independent of $m_{x}$.
	
\begin{figure}[t]
		\centering
		\includegraphics[width=1\linewidth]{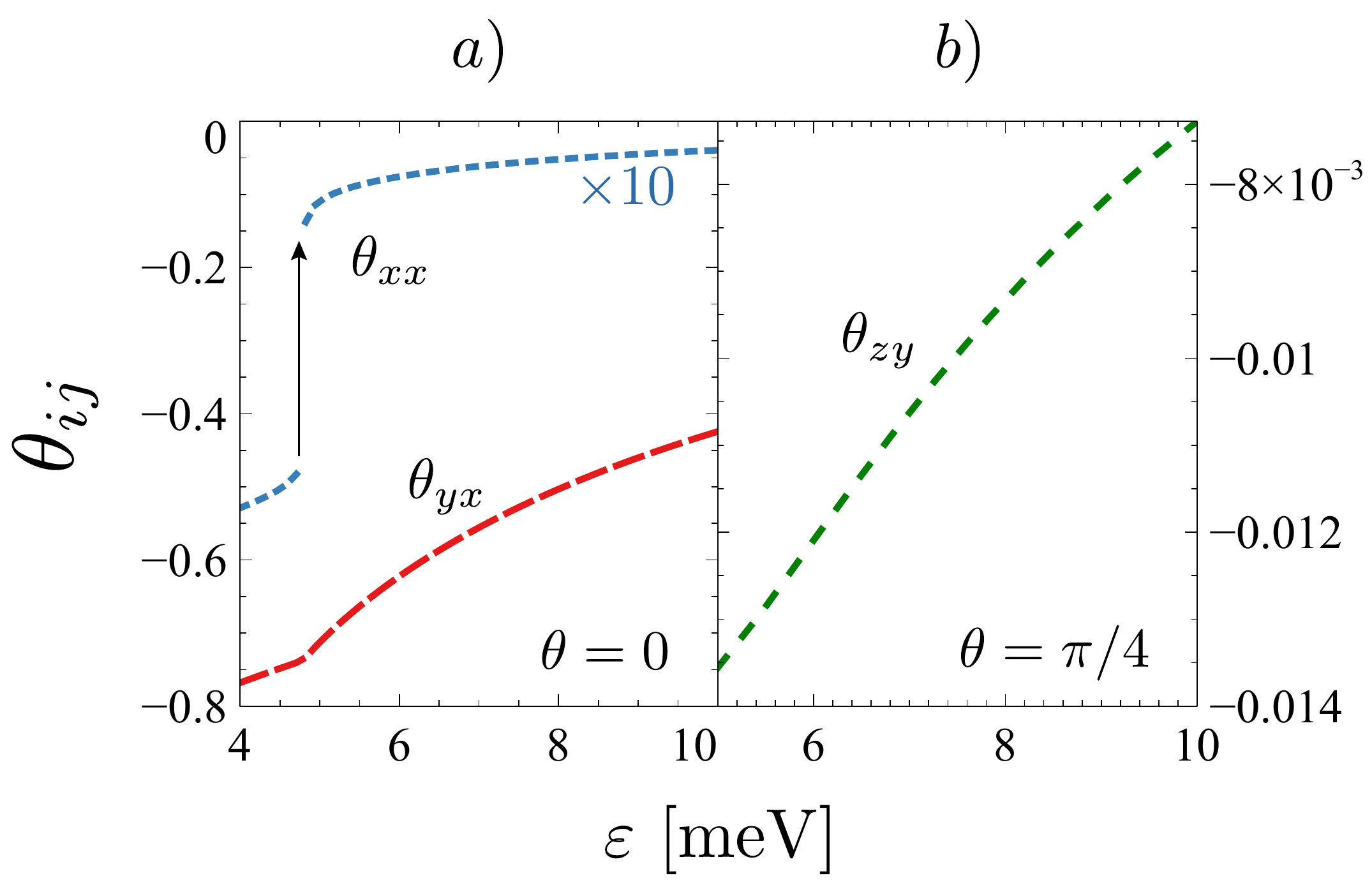}
		\caption{Current-induced torque efficiencies as functions of the Fermi energy in the strong scattering limit, with magnetization a): $\phi=0$ and b): $\phi=\pi / 4$. In the former case, the range of Fermi energies covered spans both inside and outside the spin gap, whose upper limit is $4.8$ meV, while in the latter only the regime outside the spin gap is resolved. The blue lines portrays the damping-like torque efficiency, while the red and green curves represent the field-like torque efficiencies. Parameters: $m^{*} = 0.8m_{e}$ $\alpha = 1.7 \times 10^{-11} \text{ ms}^{-1}$, $\Delta_{\text{xc}} m_{z} = 4.8$ meV, $n = 5 \times 10^{14} \text{ m}^{-2}$.}
		\label{Theta_xx_yy_yx_2DEG}
	\end{figure}

	Next we calculate the full spin susceptibility tensor and the CSC efficiency. The disorder self-energy has the form
	\begin{equation}
		\Sigma^{\pm} = \sum^{3}_{\alpha=0} n [g_{\alpha}(\varepsilon,\alpha,\mathbf{m}) \pm i\Gamma_{\alpha}(\varepsilon,\alpha,\mathbf{m})] s_{\alpha},
		\label{Disorder_self_energy_2DEG}
	\end{equation}
	where we note that the self-energy has now acquired a matrix structure (unlike in the Gaussian approximation), $g_{\alpha}$ and $\Gamma_{\alpha}$ are real functions, and $g_{2} = \Gamma_{2} = 0$. With this we acquire the disorder averaged Green's function by inserting Eq. (\ref{Disorder_self_energy_2DEG}) into Eq. (\ref{Disorder_averaged_GF_definition}), which in turn allows us to find $\hat K_{J}$ and hence the CSC efficiency. Going forward we shall work in the limit of strong SOC, i.e. $n\Gamma_{0} \ll \alpha p_{F}$. We may now start to analyze the damping-like and field-like torques. We begin by considering how the CSC efficiency of the damping-like torque, $\theta_{xx(yy)}$, depends on the Fermi energy for $m_{x} = 0$ within the strong scattering limit, which is shown in Fig. \ref{Theta_xx_yy_yx_2DEG}a. Here we see a discontinuity in $\theta_{xx(yy)}$, which can be attributed to breaching the upper limit of the spin gap, where-after the magnitude of the CSC efficiency decreases monotonically and smoothly with increasing $\varepsilon$. This efficiency reduction can be explained by noting that the difference in occupation numbers of the two Fermi rings becomes less significant by increasing the Fermi energy. In the limit of $\varepsilon \rightarrow \infty$, there is a total overlap of the two Fermi rings that provide opposing contributions to $S_{x(y)}$; the result is a zero diagonal response.
	
	For comparison we present the FL CSC efficiency, $\theta_{xy(yx)}$, also in Fig. \ref{Theta_xx_yy_yx_2DEG}a. Here we again see a monotonic decrease of the efficiency above of the spin gap, though, the efficiencies are two orders of magnitude larger than their damping-like counterparts outside of the spin gap. Consequently, the SOT is dominated by the FL-REE mechanism in this energy region. However, inside the spin gap we find a giant damping-like response with $\theta_{xx(yy)}$ approaching 7\% of the FL efficiency $\theta_{xy(yx)}$. Such a significant DL torque CSC efficiency cannot be achieved using perturbative methods, like the FBA and Gaussian approximations, which neglect skew scattering and therefore predict $K_{xx}=K_{yy}=K_{zx}=0$ (and hence $\tau^\text{e}_{1,2}=0$). Moving away from small in-plane magnetization, we find that $K_{zy}$ renormalizes the FL SOT. We present the CSC efficiency of this term for $\theta = \pi/4$ in Fig. \ref{Theta_xx_yy_yx_2DEG}b. We find that the $K_{zy}$ term can reach up to 2\% of the value of the REE-FL terms in this case.
	
	Let us now consider the weak scattering limit (WSL) where we may expand the response functions in powers of $u$. To write analytic expressions, we will need to assume a small in-plane magnetization, so we shall initially consider the regime $\Delta_{x} \ll \Delta_{z} \ll \alpha \ll \varepsilon$  and denote it by using a tilde. In this case, we expand the spin susceptibility to first order in $m_{x}$ to yield
	\begin{equation}
		\widetilde{K} =- \frac{e}{2\pi n} \begin{pmatrix}
			-\frac{(m^{*})^{2}\Delta_{z}\alpha}{2\pi\varepsilon} & \frac{\alpha}{u^{2}} \\
			-\frac{\alpha}{u^{2}} & -\frac{(m^{*})^{2}\Delta_{z}\alpha}{2\pi\varepsilon} \\
			\frac{m^{*} \Delta_{x} \Delta_{z}^{2}}{4\pi\alpha\varepsilon^{2}} & -\frac{\Delta_{x} \Delta_{z}}{2\pi u\alpha\varepsilon}
		\end{pmatrix},
		\label{2DEG_response_function1}
	\end{equation}
	where all elements are non-zero as expected. We next note that $\sigma_{xx(yy)} = e^{2}\varepsilon/(\pi nm^{*} u^{2})$ in the weak scattering limit for a large Fermi energy. Using Eq. (\ref{Torque_efficiencies_general}) and Eq. (\ref{2DEG_response_function1}) we find the following torque efficiencies,
	\begin{equation}
	\begin{split}
		\tilde{\tau}_{1}^{\text{o}} &= -\frac{m^{*}}{2e\varepsilon} \bigg( \alpha + \frac{\Delta_{x}^{2}u}{2\pi\alpha\varepsilon} \bigg), \\[2ex]
		\tilde{\tau}_{2}^{\text{o}} &= -\frac{m^{*}\Delta_{\text{xc}}^{2}u}{4\pi e\alpha\varepsilon}, \\[2ex]
		\tilde{\tau}_{1}^{\text{e}} &= \frac{(m^{*})^{3}\Delta_{\text{xc}} \alpha u^{2}}{4\pi e\varepsilon^{2}}, \\[2ex]
		\tilde{\tau}_{2}^{\text{e}} &= -\frac{(m^{*})^{2}\Delta_{\text{xc}}u^{2}}{4\pi e\varepsilon^{2}} \left( m^{*}\alpha - \frac{\Delta_{z}^{2}}{2\alpha\varepsilon} \right).
	\end{split}
	\end{equation}
	
	Equation (\ref{2DEG_response_function1}) shows that the REE terms ($\tilde K_{xy},\tilde K_{yx}$) are proportional to $u^{-2}$, and are thus captured by the typical Gaussian white noise distribution applied in the FBA. To capture the other entries we consider the next order at $u^{-1}$, which requires the prescription of a non-Gaussian average of the form $\langle V(\mathbf{x}) V(\mathbf{x}') V(\mathbf{x}'') \rangle = nR^{6}u^{3}\delta(\mathbf{x}-\mathbf{x}')\delta(\mathbf{x}'-\mathbf{x}'')$ \cite{milletari_quantum_2016}. The physics of this triple-scattering within the response function is captured by truncating the $T$-matrix series at the third order in Fig. \ref{Diagrammatics}b \cite{milletari_quantum_2016}. Finally, the damping-like elements, $\mathcal{O}(u^{0})$, can be found by calculating fourth-order scattering diagrams.
	
	An alternative regime to that above is one in which the SOC is weaker than the out-of-plane magnetization, $\Delta_{x} \ll \alpha \ll \Delta_{z} \ll \varepsilon$,  which we denote by an overline. In this case, we find
	\begin{equation}
		\bar{K} = -\frac{e}{2\pi n} \begin{pmatrix}
			-\frac{(m^{*})^{3}\alpha^{3}}{2\pi\Delta_{z}} & \frac{\alpha}{u^{2}} \\
			-\frac{\alpha}{u^{2}} & -\frac{(m^{*})^{3}\alpha^{3}}{2\pi\Delta_{z}} \\
			\frac{(m^{*})^{2} \alpha \Delta_{x}}{4\pi\epsilon} & -\frac{m^{*} \alpha \Delta_{x}}{2\pi u\Delta_{z}},
		\end{pmatrix}
		\label{2DEG_response_function2}
	\end{equation}
	which yields
	\begin{equation}
		\begin{split}
			\bar{\tau}_{1}^{\text{o}} &= -\frac{m^{*} \alpha }{2e\varepsilon} \bigg( 1 + \frac{m^{*}\Delta_{x}^{2}u}{2\pi \Delta_{z}^{2}} \bigg), \\[2ex]
			\bar{\tau}_{2}^{\text{o}} &= -\frac{(m^{*})^{2} \alpha \Delta^{2}_{\text{xc}} u}{4\pi e \Delta_{z}^{2}\varepsilon}, \\[2ex]
			\bar{\tau}_{1}^{\text{e}} &= \frac{(m^{*})^{4} \alpha^{3}\Delta_{\mathrm{xc}} u^{2}}{4\pi e \Delta_{z}^{2}\varepsilon}, \\[2ex]
			\bar{\tau}_{2}^{\text{e}} &= -\frac{\alpha (m^{*})^{3} u^{2}\Delta_{\text{xc}}}{4\pi e\varepsilon} \left( \frac{m^{*}\alpha^{2}}{\Delta_{z}^{2}}-  \frac{1}{2\varepsilon} \right).
		\end{split}
	\end{equation}
	
\begin{figure}[t]
		\centering
		\includegraphics[width=1\linewidth]{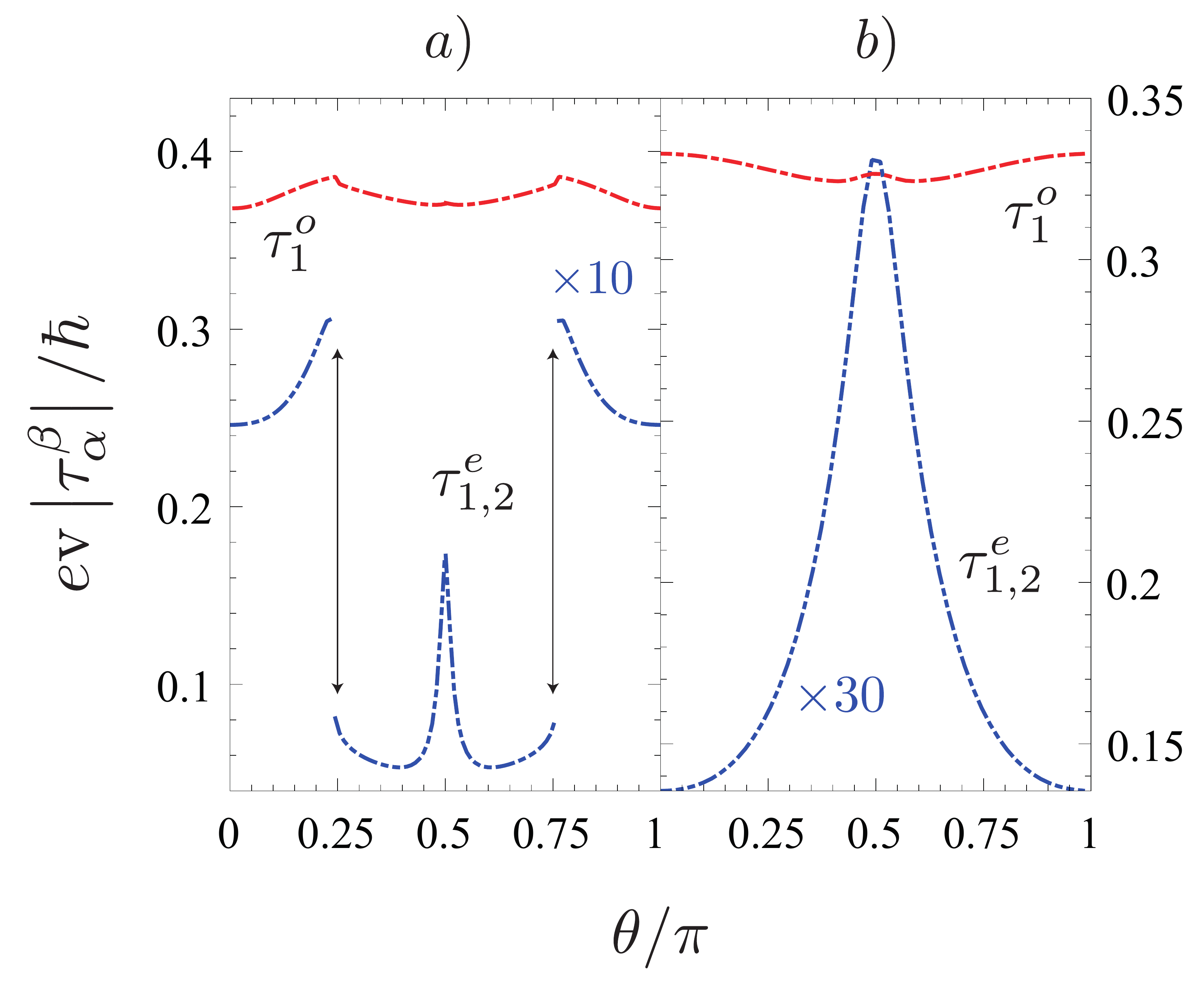}
		\caption{Torque efficiencies as a function of magnetization angle, with periodicity $\pi$. The field-like torque efficiency $\tau_{1}^{\text{o}}$ is denoted by the red line, whilst the damping-like efficiencies are equal in module, $\left| \tau_{1}^{\text{e}} \right|= \left|\tau_{2}^{\text{e}} \right|$, and represented by the blue line. In a) the Fermi energy is inside the spin gap for $\theta=0$, i.e., $\varepsilon=4.5\,\mathrm{meV}$. The discontinuous behaviour reflects the transition in the electronic band structure from inside to outside the spin gap. In b), the Fermi energy is above the spin gap, $\varepsilon=5.5\,\mathrm{meV}$, and the torque coefficients are smooth functions of $\theta$.
		Parameters $m^{*} = 0.8\,m_{e}$ (Tantalum), $v = 5 \times 10^{4} \,\text{m/s}$, $\Delta_{\text{xc}} = 5$ meV, and $n = 0.5 \times 10 ^{15}\, \text{m}^{-1}$.} 
		\label{Angle_DEP}
	\end{figure}	
	
Next, we present the dominant FL and DL SOTs in Fig. \ref{Angle_DEP} as functions of the magnetization angle $\theta$ in the strong scattering limit. The left panel (Fig. \ref{Angle_DEP}a) considers a Fermi energy inside the spin gap for an initial out-of-plane configuration of the magnetization. This is the \textit{strong damping} regime, where the ratio between the DL and FL torques is maximized (see Fig. \ref{Theta_xx_yy_yx_2DEG}). As the FM's magnetization is shifted from purely out-of-plane to purely in-plane (i.e. $\theta = 0 \rightarrow \pi/2$), the spin gap between the bands begins to shrink and vanishes when $\theta = \pi/2$. Consequently, the fixed Fermi energy will only intersect a single band for smaller angles, before then intersecting both bands at some \textit{critical angle}, $\theta_{c}(\varepsilon)$ ($\sim \pi/4$ in this case), where the spin gap has shrunk sufficiently to allow this, and hence a discontinuity is observed at this $\theta_{c}(\varepsilon)$. This corresponds to moving from the strong damping regime to the \textit{weak damping regime}. The angular dependence of the SOT coefficients is clearly symmetric about $\theta = \pi/2$. On the other hand, if the Fermi energy is instead situated above the spin gap at $\theta = 0$, it will remain outside the spin gap for all magnetization angles and hence the system will always be in the weak damping regime. Therefore, the torque coefficients will be smooth continuous functions of $\theta$, see Fig. \ref{Angle_DEP}b. This complete angular description of $\tau^{o}_{1}$ and $\tau^{e}_{1,2}$ for strong disorder is enabled by treating the impurity potential and the magnetic exchange coupling on equal footing (i.e. a full $T$-matrix numerical treatment with a generic $\Delta_{\text{xc}}$).

 We can also see that in both cases of Fig. \ref{Angle_DEP} that the standard FL contribution has a relatively weak angular dependence. Hence, $\tau_{1}^{\text{o}}$ may be treated as approximately constant to first approximation inline with previous literature \cite{ado_microscopic_2017}. In contrast, the DL torque coefficients, which are controlled entirely by the non-perturbative $K_{xx(yy)}$ components, exhibit a strong dependence upon the magnetization angle; a dependence that would otherwise be missed in perturbative methods. Clearly, the approximation of the DL torque coefficients as constants therefore breaks down.

\section{Conclusions} \label{Sec_conclusions}

We have demonstrated that a complete understanding of damping-like torques in diffusive NM/FM bilayers hinges on scattering processes beyond the Gaussian approximation. Specifically, we showed that skew scattering is essential for a correct description of interfacial SOT already in diffusive systems characterized by impurities with weak scattering potentials. By treating both the disorder potential and spin-dependent interactions in the NM band structure non-perturbatively, we have gained access to SOTs in the strong scattering limit. Here we found that ultra-thin NM/FM bilayers host an efficient skew scattering-activated damping-like SOT generated purely at the interface. This indicates that the bulk contribution to SOT (the SHE) is not a necessity for inducing magnetic switching of the FM, which may help shed light on recent experiments on ultra-thin NM/FM bilayers which observed SOT driven switching of the FM's magnetization \cite{kim_layer_2013}.

As another application of the nonperturbative approach introduced here, we showed that the DL torque exhibited a non-trivial angular dependence upon the magnetization, thus illustrating the limitation of assuming the proximity-induced Zeeman coupling in the NM Hamiltonian to be completely out of plane (a common approximation in the literature). When the Fermi energy was located above the spin gap we saw a dramatic increase in the DL torque as the out-of-plane magnetization approached zero. Similarly, for a Fermi energy inside the spin gap, we also observed a rapid increase in the DL torque whilst the FL torque remained approximately constant. However, at some magnetization angle a discontinuity in both the DL and FL torques was encountered due to the shrinking of the spin gap.

Given the scientific and technological importance of interfacial SOTs, the investigation of other exotic materials (e.g. topological insulators and Weyl semimetals) is of significant interest. What makes these materials so interesting is their unusual electronic structure and naturally strong SOC, which may give rise to interesting spin-charge inter-conversion processes. The formulation of a nonperturbative SOT theory for those systems could unlock yet new manifestations of higher-order scattering processes.

%

	
\end{document}